\documentclass{collective-intelligence}

\usepackage{graphicx}
\usepackage{subfigure}
\usepackage{bbm}
\usepackage{amsmath}
\usepackage{wrapfig}
\usepackage{array}

\newcommand{\indicator}[1]{\mathbbm{1}_{\left[ {#1} \right] }}

\begin{document}
\bibliographystyle{agsm}

\title{LEARNING TO PREDICT THE WISDOM OF CROWDS}
%
%
%
%
%

\numberofauthors{1} 
%
\author{
%
%
\alignauthor ~~Seyda Ertekin${^{1,3}}$\qquad\qquad~~~ Haym Hirsh${^{2,3}}$\qquad\qquad~~~ Cynthia Rudin${^{1,3}}$\\ 
              \email{~~seyda@mit.edu \qquad ~~~~~~ hirsh@cs.rutgers.edu \qquad~~~~~~~ rudin@mit.edu~~~~~~}
 \and
\alignauthor 
       \affaddr{$^1$MIT Sloan School of Management, Massachusetts Institute of Technology, Cambridge, MA 02142}\\
       \affaddr{$^2$Department of Computer Science, Rutgers University, Piscataway, NJ 08854}\\
       \affaddr{$^3$MIT Center for Collective Intelligence, Massachusetts Institute of Technology, Cambridge, MA 02142}\\
}
\maketitle
\begin{abstract}
The problem of ``approximating the crowd" is that of estimating the crowd's majority opinion by querying only a subset of it. Algorithms that approximate the crowd can intelligently stretch a limited budget for a crowdsourcing task. We present an algorithm, ``CrowdSense," that works in an online fashion to dynamically sample subsets of labelers based on an exploration/exploitation criterion. The algorithm produces a weighted combination of a subset of the labelers' votes that approximates the crowd's opinion.
\end{abstract}

\vspace{-1mm}
\section{Introduction}
Crowdsourcing systems are a useful way to obtain many opinions (or ``votes") very quickly. However, in cases where each vote is provided at a cost, collecting a vote from every member of the crowd in order to determine the majority opinion can be expensive and may not even be attainable under fixed budget constraints. Because of the open nature of crowdsourcing systems, it is not necessarily easy to approximate the majority vote of a crowd on a budget by sampling a representative subset of the voters. For example, the crowd may be comprised of labelers with a range of capabilities, motives, knowledge, views, personalities, etc. Without any prior information about the characteristics of the labelers, a small sample of votes is not guaranteed to align with the true majority opinion. In order to effectively approximate the crowd, we need to determine who are the most representative members of the crowd, in that they can best represent the interests of the crowd majority. This is even more difficult to accomplish when items arrive over time, and it requires our budget to be used both for 1)~estimating the majority vote, even before we understand the various qualities of each labeler, and 2)~exploring the various labelers until we can estimate their qualities well. 

Estimating the majority vote in particular can be expensive before the labelers' qualities are known, and we do not want to pay for additional votes that are not likely to impact the decision. In order to make economical use of our budget, we could determine when just enough votes have been gathered to confidently align our decision with the crowd majority.  The budget limitation necessitates a compromise: if we pay for a lot of votes per decision, our estimates will closely align with the crowd majority, but we will only make a smaller number of decisions, whereas if we pay for less votes per decision, the accuracy may suffer, but more decisions are made. There is clearly an exploration/exploitation tradeoff: before we can exploit by using mainly the best labelers, we need to explore to determine who these labelers are, based on their agreement with the crowd.

The main contributions of this work are a modular template for algorithms that approximate the wisdom of the crowd, including the exploitation/exploration choices, and an algorithm, \textit{CrowdSense}, that arises from specific choices within this template. In an online fashion, CrowdSense dynamically samples a subset of labelers, determines whether it has enough votes to make a decision, requests more if the decision is sufficiently uncertain, and iteratively updates the labelers' quality estimates. CrowdSense keeps a balance between exploration and exploitation in its online iterations -- exploitation in terms of seeking labels from the highest-rated labelers, and exploration so that enough data are obtained about each labeler to ensure that we learn each labeler's accuracy sufficiently well.  Our experimental results demonstrate the effectiveness of CrowdSense for identifying the representative members of the crowd and approximating the crowd majority. In our discussions, a representative labeler is a member of the crowd whose votes are consistent with the crowd's vote most of the time, and hence their votes are highly informative for accurately approximating the crowd's majority vote. Throughout the paper, a ``majority vote" refers to the simple, every day sense of voting wherein every vote is equal, with no differential weighting of the votes. This is in contrast to a weighted majority vote, as we use in our algorithm, wherein each labeler's vote is multiplied by the labeler's quality estimate. This weighting scheme ensures that the algorithm places a higher emphasis on the votes of higher quality labelers. 

The rest of the paper is organized as follows: The next section presents a discussion on related work. The following section provides details on CrowdSense within the general context of a modular framework for approximating the wisdom of the crowd. We then present the outline of the datasets that we used and the baselines we considered, followed by experiments that demonstrate the performance of CrowdSense against the baselines, and present the impact of specific choices within the general framework. The last section presents concluding remarks and directions for future work.

\vspace{-1mm}
\section{Related Work}
The low cost of crowdsourcing labor has increasingly led to the use of resources such as Amazon Mechanical Turk\footnote{http://www.mturk.com} (AMT) to label data for machine learning purposes, where collecting multiple labels from non-expert annotators can yield results that rival those of experts. This cost-effective way of generating labeled collections using AMT has also been used in several studies \cite{Dakka_2008,Kaisser_2008,Nakov_2008,Snow_2008,Sorokin_2008,Nowak_2010}. While crowdsourcing clearly is highly effective for easy tasks that require little to no training of the labelers, the rate of disagreement has been shown to increase with task difficulty \cite{Gillick_2010,Sorokin_2008}. 
Although a range of approaches are being developed to manage the varying reliability of crowdsourced labor \citeaffixed{Callison-Burch_2010,Law_2011,Quinn_2011,Wallace_2011}{see, for example}, the most common method for using crowdsourcing to label data is to obtain multiple labels for each item from different labelers and treat the majority label as an item's true label. For example, \citeasnoun{Sheng_2008} demonstrated that repeated labeling can be preferable to single labeling in the presence of label noise, especially when the cost of data preprocessing is non-negligible. \citeasnoun{Dekel_2009b} proposed an algorithm for pruning the labels of less reliable labelers in order to improve the accuracy of the majority vote of labelers. First collecting labels from labelers and then discarding the lower quality ones presents a different viewpoint than our work, where we achieve the same ``pruning effect" by estimating the qualities of the labelers and not asking the low quality ones to vote in the first place. A number of researchers have explored approaches that learn how much to trust different labelers, typically by comparing each labeler's predictions to the majority-vote prediction of the full set.  These approaches often use methods to learn both labeler quality characteristics and latent variables representing the ``true" labels of items \citeaffixed*{Dawid_1979,Kasneci_2011,Smyth_1994b,Smyth_1994a,Warfield_2004}{e.g.}, sometimes in tandem with learning values for other latent variables such as task difficulty \cite{Welinder_2010a,Whitehill_2009}, classifier parameters \cite{Dekel_2009,Raykar_2010,Yan_2010b,Yan_2010a}, or domain-specific information about the labeling task \cite{Welinder_2010a}. 

Our work appears similar to the preceding efforts in that we similarly seek predictions from multiple labelers on a collection of items, and seek to understand how to assign weights to them based on their prediction quality. However, previous work on this topic viewed labelers mainly as a resource to use in order to lower uncertainty about the true labels of the data. 
In this paper, we seek to approximate the correct prediction at lower cost by decreasing the number of labelers used, as opposed to increasing accuracy by turning to additional labelers at additional cost. In other words, usually the classifier is not known and the task is to try to learn it, whereas here the classifier is known and the task is to approximate it. This work further differs from most of the preceding efforts in that they presume that learning takes place after obtaining a collection of data, whereas our method works in online settings, where it simultaneously processes a stream of arriving data while learning the different quality estimates for the labelers. \citeasnoun{Sheng_2008} is one exception, performing active learning by reasoning about the value of seeking additional labels on data given the data obtained thus far. \citeasnoun{Donmez_2009} simultaneously estimate labeler accuracies and train a classifier using labelers' votes to apply active learning to select the next example for labeling. We discuss the approach taken by \citeasnoun{Donmez_2009} later in the paper as one of the baselines to which we compare our results, because this is the only algorithm we know of aside from ours that can be naturally applied to approximating the crowd in the online setting.
 
Finally, our work appears superficially similar to what is accomplished by polling processes.  Polls obtain the opinions of a small number of people so as to approximate the opinions of a typically much large population.  However, polling presumes knowledge of individuals' demographic characteristics, determining how to extrapolate from the views of a small group with certain demographic characteristics to a desired target population with its own demographic characteristics.  Our work knows nothing about the demographics of the labelers, and at its core assumes that however closely a labeler matches the overall crowd is a good predictor of whether the labeler will do so in the future. Incorporating knowledge of labeler demographics, in order to predict the qualities of labelers, is an intriguing direction for future work.

\vspace{-1mm}
\section{CrowdSense}
\label{sec:algo}
Let us first model the labelers' quality estimates as a measure of their agreement with the crowd. Let $L=\{l_1, l_2,\ldots, l_M\},~ l_k:x\rightarrow\{-1,1\}$ denote the set of labelers and $\{x_1, x_2,\ldots, x_t,\ldots,x_N\}$ denote the sequence of examples, which could arrive one at a time. We define $V_{it}:=l_i(x_t)$ as $l_i$'s vote on $x_t$ and $S_t \subset \{1,\ldots,M\}$ as the set of labelers selected to label $x_t$.  For each labeler $l_i$, we then define $c_{it}$ as the number of times we have observed a label from $l_i$ so far:
\begin{equation}
c_{it} := \sum_{\tilde{t}=1}^{t}\indicator{i \in S_{\tilde{t}}}
\end{equation}
and define $a_{it}$ as how many of those labels were consistent with the other labelers:
\begin{equation}
 a_{it} := \sum_{\tilde{t}=1}^{t}\indicator{i \in S_{\tilde{t}}, V_{i\tilde{t}} =V_{S_{\tilde{t}}\tilde{t}}}
\end{equation}
where $V_{S_{t}{t}} = \textrm{sign}\left(\sum_{i \in S_{t}}V_{it}Q_{it}\right)$ is the weighted majority vote of the labelers in $S_t$. Labeler $l_i$'s quality estimate is then defined as
\begin{equation}
\label{eq:labeler_quality}
Q_{it} = \frac{a_{it} + K}{c_{it} + 2K}
\end{equation}
where $t$ is the number of examples that we have collected labels for and where $K$ is a smoothing parameter. $Q_{it}$ is a Bayesian shrinkage estimate of the probability that labeler $i$ will agree with the crowd, pulling values down toward $1/2$ when there are not enough data to get a more accurate estimate. This ensures that labelers who have seen fewer examples are not considered more valuable than labelers who have seen more examples and whose performance is more certain. Labelers who are accurate less than half the time can be ``flipped" so the opposite of their votes are used.

\begin{figure}[b!]
\begin{enumerate}
  
  \item \textbf{Input:} Examples $\{x_1,x_2,\ldots,x_N\}$, Labelers $\{l_1,l_2,\ldots,l_M\}$, confidence threshold $\varepsilon$, smoothing parameter $K$.

  \item \textbf{Define:} $L_Q = \{l^{(1)},\ldots, l^{(M)}\}$, labeler id's in descending order of their quality estimates.

  \item  \textbf{Initialize:} $a_{i1} \leftarrow 0,~c_{i1} \leftarrow 0 \textrm{ for } i=1,\ldots,M$. Initialize $Q_{it} \leftarrow 0 \textrm{ for } i=1\ldots M, t = 1\ldots N$

  \item  \textbf{Loop for} $t=1,...,N$
  \begin{enumerate}
     \item Compute quality estimates $Q_{it} = \frac{a_{it} + K}{c_{it} + 2K},~i=1,\ldots,M$. Update $L_Q$.
     \item $S_t = \{l^{(1)}, l^{(2)}, k\}$, where $k$ is randomly sampled from the set $\{3,\ldots, M\}$.
     \item \textbf{Loop for} $j=3\ldots M,~j\neq k$
     \begin{enumerate}
         \item $\textrm{Score}(S_{t}) = \sum_{i \in S_{t}} V_{it}Q_{it},~~l_{\textrm{\small candidate}} = l^{(j)}$.
	 \item If $\frac{\left|\textrm{Score}(S_{t})\right| - Q_{l_\textrm{\tiny{candidate}},t}}{\left|S_{t}\right|+1} < \varepsilon$, then $S_t \leftarrow S_t \cup {l_{\textrm{candidate}}}$. Otherwise exit loop to stop adding new labelers to $S_t$.
     \end{enumerate}
     \item Get the weighted majority vote of the labelers $V_{S_{t}{t}} = \textrm{sign}\left(\sum_{i \in S_{t}}V_{it}Q_{it}\right)$
     \item $\forall i \in S_{t} \textrm{ where } V_{it} =V_{S_{t}t},~~ a_{it} \leftarrow a_{it}+1$
     \item $\forall i \in S_{t},~~ c_{it} \leftarrow c_{it}+1$
  \end{enumerate}
    \item \textbf{End}
\end{enumerate}
\caption{{Pseudocode for the CrowdSense algorithm.}
\label{fig:pseudo_crowdsense}}
\end{figure}

Pseudocode for the CrowdSense algorithm is given in Figure \ref{fig:pseudo_crowdsense}. At the beginning of an online iteration to label a new example, the labeler pool is initialized with three labelers; we select two labelers that have the highest quality estimates $Q_{it}$ and select another one uniformly at random. This initial pool of seed labelers enables the algorithm to maintain a balance between exploitation of quality estimates and exploration of the quality of the entire set of labelers. We ask each labeler to vote on the example, and we pay a fixed price per label. The votes obtained from these labelers for this example are then used to generate a \textit{confidence score}, given as 
\begin{equation*}
\textrm{Score}(S_{t}) = \sum_{i \in S_{t}} V_{it}Q_{it}
\end{equation*}
which represents the weighted majority vote of the labelers. Next, we determine whether we are certain that the sign of $\textrm{Score}(S_{t})$ reflects the crowd's majority vote, and if we are not sure, we repeatedly ask another labeler to vote on this example until we obtain sufficient certainty about the label. To measure how certain we are, we look at the value of $\left|\textrm{Score}(S_{t})\right|$ and select the labeler with the highest quality estimate $Q_{it}$ that is not in $S_{t}$ as a candidate to label this example. We then check whether this labeler could potentially either change the weighted majority vote if his vote were included, or if his vote could bring us into the \textit{regime of uncertainty} where the $\textrm{Score}(S_{t})$ is close to zero, and the vote is approximately a tie. The criteria for adding the candidate labeler to $S_{t}$ is defined as:
\begin{equation}
\frac{\left|\textrm{Score}(S_{t})\right| - Q_{l_\textrm{\tiny{candidate}},t}}{\left|S_{t}\right|+1} < \varepsilon
\label{eq:add_criteria}
\end{equation}
where $\varepsilon$ controls the level of uncertainty we are willing to permit, $0 < \varepsilon \leq 1$. If (\ref{eq:add_criteria}) is true, the candidate labeler is added to $S_{t}$ and we get this labeler's vote for $x_t$. We then recompute $\textrm{Score}(S_{t})$ and follow the same steps for the next-highest-quality candidate from the pool of unselected labelers. If the candidate labeler is not added to $S_{t}$, we assign the weighted majority vote of the current sub-crowd, $S_t$, as the predicted label of this example and proceed to label the next example in the collection. 

\begin{table*}[t!]
\begin{minipage}[b]{0.08\linewidth}\centering
\begin{tabular}{l}
\multicolumn{1}{r}{} \\
{\bf MovieLens} \\
{\bf ChemIR} \\
{\bf Reuters} \\
{\bf Adult} \\
\end{tabular}
\end{minipage}
\hspace{0.5cm}
\begin{minipage}[b]{0.1\linewidth}\centering
\begin{tabular}{|c|} \hline
\# Examples \\ \hline \hline
137 \\ \hline
1,165 \\ \hline
6,904 \\ \hline
32,561 \\ \hline
\end{tabular}
\end{minipage}
\hspace{0.5cm}
\begin{minipage}[b]{0.8\linewidth}
\setlength{\tabcolsep}{3.4pt}
\begin{tabular}{|c c c c c c c c c c c c  c |} \hline
 L1 & L2 & L3 & L4 & L5 & L6 & L7 & L8 & L9 & L10 & L11 & L12 &  L13\\ \hline \hline
 48.17 & 89.78 & 93.43 & 48.90 & 59.12 & 96.35 & 87.59 & 54.01 & 47.44 & 94.16 & 95.62 & --& --\\ \hline
 50.72 & 46.78 & 84.46 & 88.41 & 86.69 & 87.46 & 49.52 & 78.62 & 82.06 & 50.12 & 50.98& --& --\\ \hline
 80.76 & 83.00 & 89.70 & 82.98 & 88.12 & 87.04 & 95.42 & 80.21 & 78.68 & 95.06 & 82.88 & 71.57 & 87.54\\ \hline
 81.22 & 80.59 & 86.22 & 87.63 & 91.12 & 94.11 & 56.68 & 85.51 & 81.32 & 85.54 & 79.74 & 84.86 & 96.71\\ \hline
\end{tabular}
\end{minipage}
\caption{The number of examples in each dataset and the true accuracies of the labelers.}
\end{table*}

\section{Datasets and Baselines}
\label{sec:datasets_and_baselines}
We conducted experiments on four separate datasets that model a crowd from two separate perspectives. MovieLens is a movie recommendation dataset of user ratings on a collection of movies, and the goal is to find the majority vote of these reviewers. The dataset is originally very sparse, meaning that only a small subset of users have rated each movie. We compiled a smaller subset of this dataset where each movie is rated by each user in the subset, to enable comparative experiments. We mapped the original rating scale of [0-5] to votes of \{-1,1\} by using 2.5 as the decision boundary. ChemIR is a dataset of chemical patent documents from the 2009 TREC Chemistry Track. This track defines a ``Prior Art Search" task, where the competition is to develop algorithms that, for a given set of patents, retrieve other patents that they consider relevant to those patents. The evaluation criteria is based on whether there is an overlap of the original citations of patents and the patents retrieved by the algorithm. The ChemIR dataset that we compiled is the complete list of citations of several chemistry patents, and the \{+1,-1\} votes indicate whether or not an algorithm has successfully retrieved a true citation of a patent. Both MovieLens and ChemIR datasets have 11 labelers in total. Reuters is a popular dataset of articles that appeared on the Reuters newswire in 1987. We selected documents from the \textit{money-fx} category. The Reuters data is divided into a ``training set" and a ``test set," which is not the format we need to test our algorithms. We used the first half of the training set (3,885 examples) to develop our labelers. Specifically, we trained several machine learning algorithms on these data: AdaBoost, Na\"{i}ve Bayes, SVM, Decision Trees, and Logistic Regression, where we used several different parameter settings for SVM and Decision Trees. Each algorithm with its specific parameter setting was used to generate one labeler, and there were 10 labelers generated this way. Additionally, we selected 3 features of the dataset as labelers, for a total of 13 labelers. We combined the other half of the training set with the test set, which provided 6,904 total examples over which we used to measure the performance of CrowdSense and the baselines. The same simulation of a crowd that we conducted for the Reuters dataset was also used for the Adult dataset from the UCI Machine Learning Repository, which is collected from the 1994 Census database.  

For MovieLens, we added 50\% noise and for ChemIR we added 60\% noise to 5 of the labelers to introduce a greater diversity of judgements, since all the original labelers had comparable qualities and did not strongly reflect the diversity of labelers and other issues that we aim to address. For the Reuters and Adult datasets, we varied the parameters of the classification algorithm ``labelers" to yield predictions with varying peformance. All reported results are averages of 100 runs, each with a random ordering of examples to prevent bias due to the order in which examples are presented.

We compared CrowdSense with several baselines: (a) the accuracy of the \textit{average} labeler, represented as the mean accuracy of the individual labelers, (b) the accuracy of the overall best labeler in hindsight, and (c) the algorithm that selects just over half the labelers (\textit{i.e.} $\lceil 11/2 \rceil=6$ for ChemIR and MovieLens, $\lceil 13/2 \rceil=7$ for Reuters and Adult) uniformly at random, which combines the votes of labelers with no quality assessment using majority vote. As for the comparative analysis of labeler selection based on quality estimates,  we compare CrowdSense against IEThresh \cite{Donmez_2009}. IEThresh builds upon Interval Estimation (IE) learning, which estimates an upper confidence interval UI for the mean reward for an action, which is a technique used in reinforcement learning. In IEThresh, an action refers to asking a labeler to vote on an item, and a reward represents the labeler's agreement with the majority vote. The UI metric for IEThresh is defined for a sample ``a" as
\begin{equation}
\label{eq:iethresh}
UI(a) = m(a) + t^{(n-1)}_{\frac{\alpha}{2}}\frac{s(a)}{\sqrt{n}}
\end{equation}
where $m(a)$ and $s(a)$ are the sample mean and sample standard deviation for $a$, $n$ is the sample size and $t^{(n-1)}_{\frac{\alpha}{2}}$ is the critical value for the Student's t-distribution. The sample ``a" for a labeler is the vector of $\pm1$'s, indicating agreement of that labeler with the majority. IEThresh updates the $UI$ scores of the labelers after observing new votes, and given the $UI$ scores $\{UI_1,UI_2,\ldots,UI_k\}$ for the labelers, IEThresh selects all labelers $j$ with $UI_{j} > \varepsilon \times \max_{\tilde{j}}(UI_{\tilde{j}})$. The $\varepsilon$ parameter in both CrowdSense and IEThresh algorithms tunes the size of the subset of labelers selected for voting, so we report results for a range of $\varepsilon$ values. Note that tuning $\varepsilon$ exhibits opposite behavior in CrowdSense and IEThresh; increasing $\varepsilon$ relaxes CrowdSense's selection criteria to ask for votes from more labelers, whereas larger $\varepsilon$ causes IEThresh to have a more strict selection policy. So a given value of $\varepsilon$ for CrowdSense does not directly correspond to a particular value of $\varepsilon$ for IEThresh. On the other hand, since $\varepsilon$ controls the number of labelers used for each example in both algorithms, it also controls the total number of labelers used for the entire collection of examples. (This is proportional to the cost of the full experiment.) When we adjust the $\varepsilon$ values for CrowdSense and IEThresh so that the total number of labelers is similar, we can directly see which algorithm is more accurate, given that the same total cost is spent on each.

\begin{figure*}[t]
\centering
\hspace{-7mm}
\subfigure {
\includegraphics[width=0.265\linewidth]{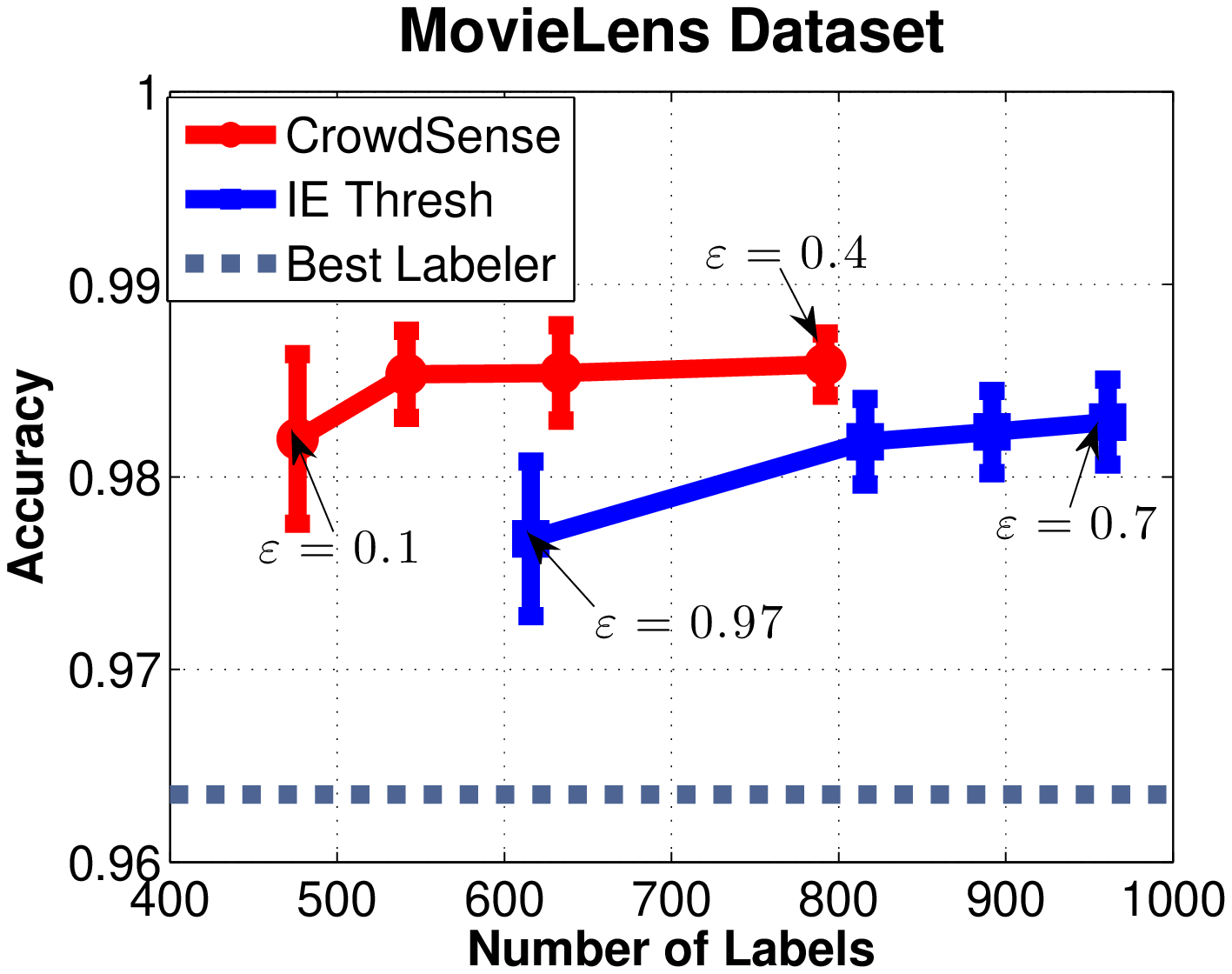}
}
\hspace{-7mm}
\subfigure {
\includegraphics[width=0.265\linewidth]{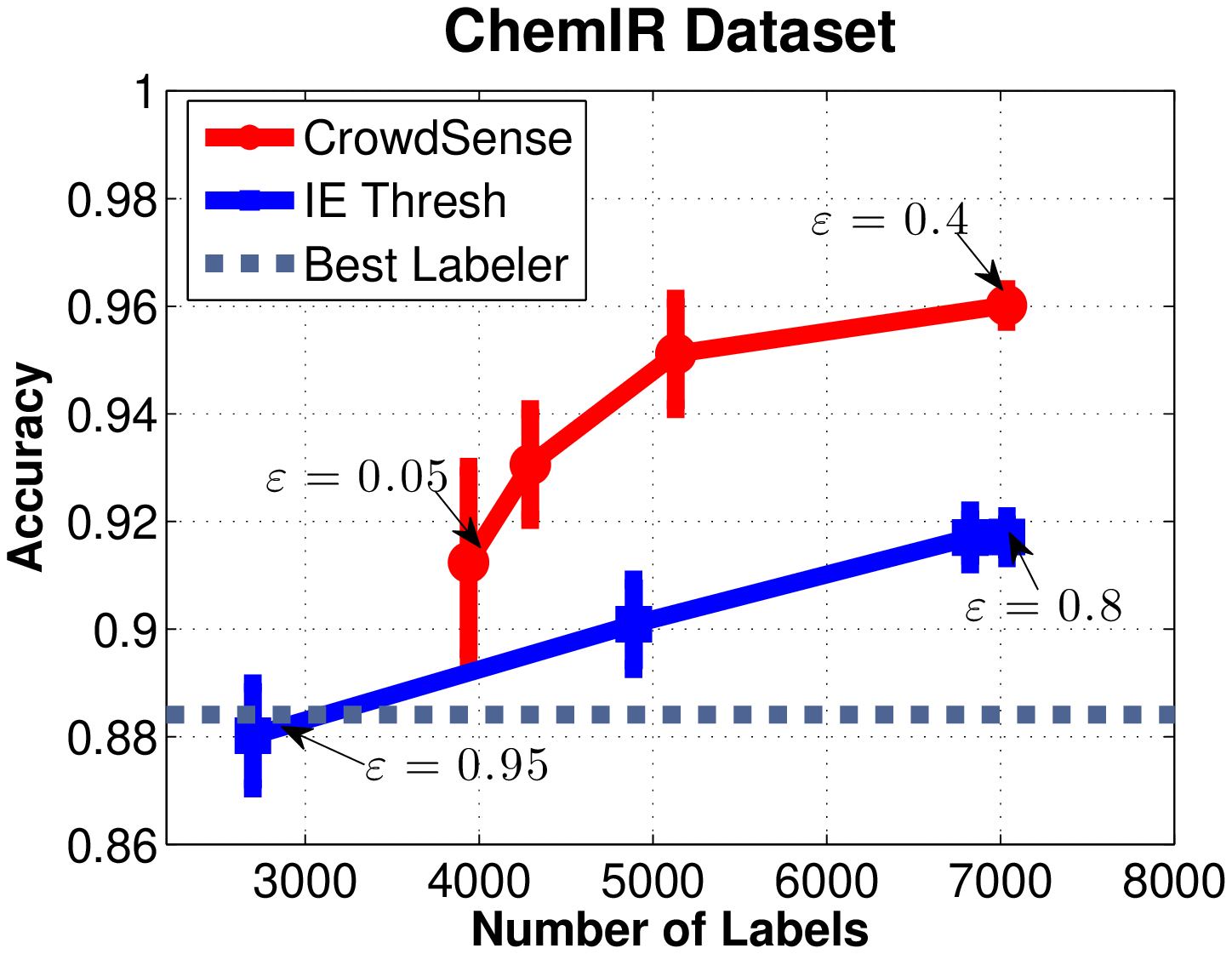}
}
\hspace{-7mm}
\subfigure {
\includegraphics[width=0.265\linewidth]{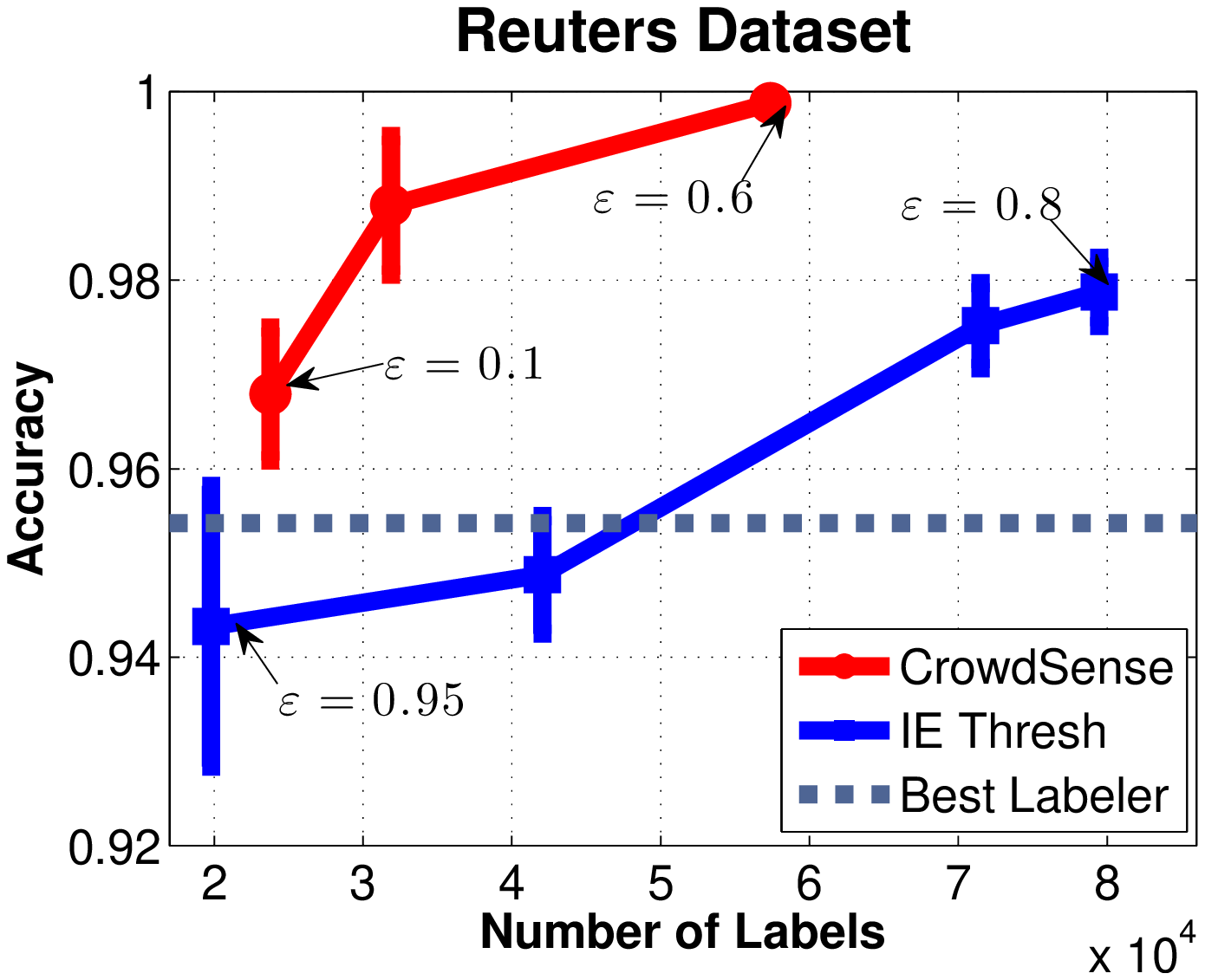}
}
\hspace{-7mm}
\subfigure {
\includegraphics[width=0.265\linewidth]{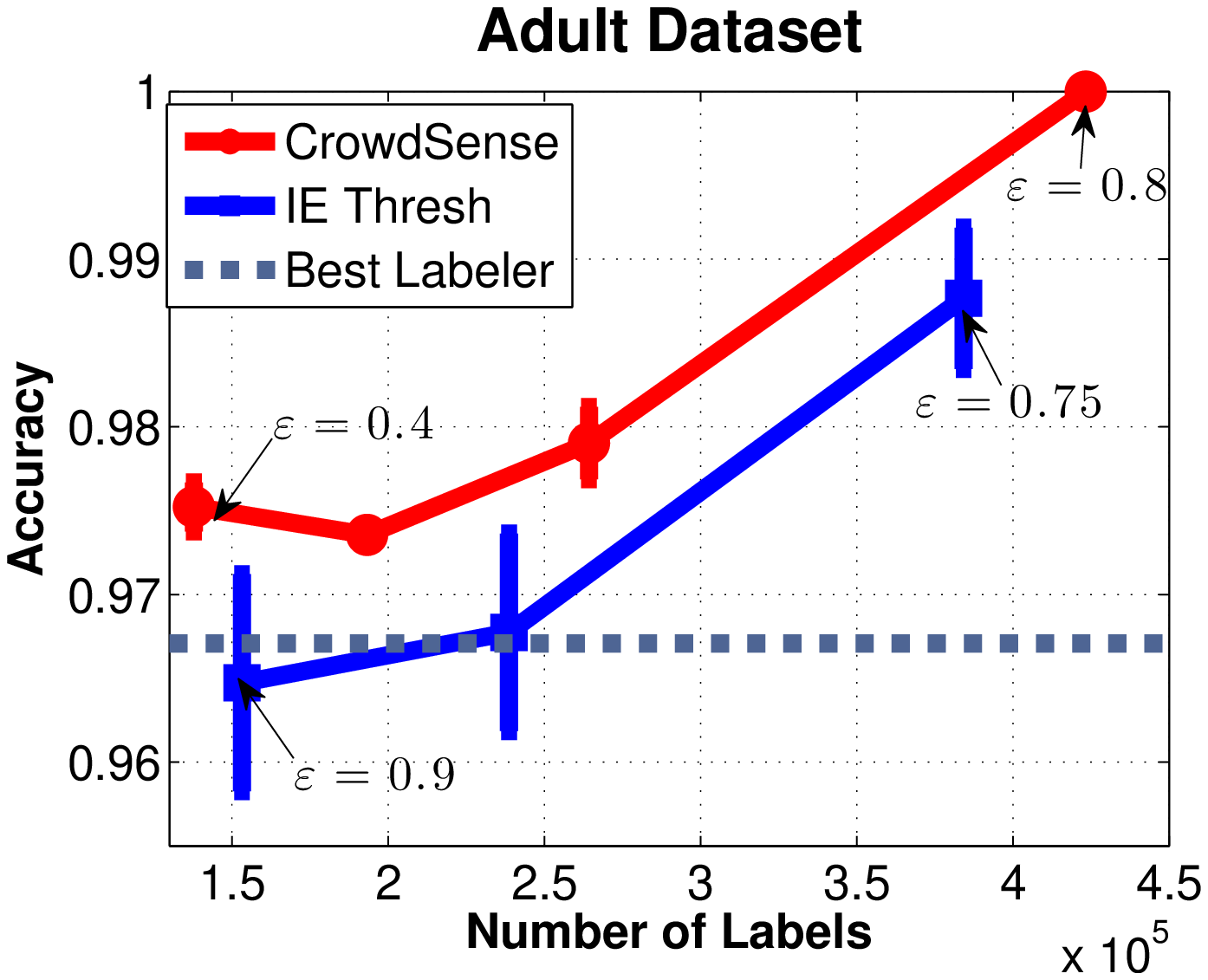}
}
\caption{Tradeoff curves, averaged over 100 runs. The x-axis is the total number of votes (the total cost) used by the algorithm to label the entire dataset. The y-axis indicates the accuracy on the full dataset. The error bars indicate sample standard deviation over the 100 runs.}
\label{fig:tradeoff}
\end{figure*}

\section{Overall Performance}
\label{sec:performance}
First, we compare CrowdSense with the baselines to demonstrate its ability to accurately approximate the crowd's vote. In the next section, we present a modular view of CrowdSense, and show the impact of each module on the algorithm's accuracy. Accuracy is calculated as the proportion of examples where the algorithm agreed with the majority vote of the entire crowd. 

Figure \ref{fig:tradeoff} indicates that uniformly across different values of $\varepsilon$, CrowdSense consistently achi-eved the highest accuracy against the baselines, indicating that CrowdSense uses any fixed budget more effectively than IEThresh. Generally, the quality estimates of CrowdSense better reflect the true accuracy of the members of the crowd and therefore, it can identify and pick a more representative subset of the crowd. The other baselines that we considered did not achieve the same level of performance as CrowdSense and IEThresh. On the subplots in Figure \ref{fig:tradeoff}, the accuracy of the best labeler in hindsight (baseline (b)) is indicated as a straight line.  Note that the accuracy of the best labeler is computed separately as a constant, and does not vary with the x-axis on the plot. Baselines (a) and (c), which are the average labeler and the unweighted random labelers, achieved performance beneath that of the best labeler. For the MovieLens dataset, the values for these baselines are 74.05\% and 83.69\% respectively; for ChemIR these values are 68.71\% and 73.13\%, for Reuters, the values are 84.84\% and  95.25\%, and for Adult they are 83.94\% and 95.03\%. The results demonstrate that asking for labels from labelers at random may yield poor performance for representing the majority vote, highlighting the importance of making informed decisions for selecting the representative members of the crowd. Asking the best and average labeler were also not effective approximators of the majority vote of the crowd. 

\section{Specific Choices for Exploration and \\Exploitation in CrowdSense}
The algorithm template underlying CrowdSense has three components that can be instantiated in different ways: (1) the composition of the initial seed set of labelers (step 4(b) in the pseudocode), (2) how subsequent labelers are added to the set (step 4(c)), and (3) the weighting scheme, which affects the selection of the initial labeler set, the way the additional labelers are incorporated, as well as the strategy for combining the votes of the labelers (steps 4(b)(c)(d)). 

We tested the effect of the first component by running separate experiments that initialized the labeler set with three (3Q), one (1Q) and no (0Q) labelers that had the highest quality estimates, where for the latter two, additional labelers were selected at random to complete the set of three initial labelers. 3Q removes the exploration capability of the initial set whereas the latter two make limited use of the quality estimates. 
\begin{figure}[h!]
\centering
\includegraphics[width=0.65\linewidth]{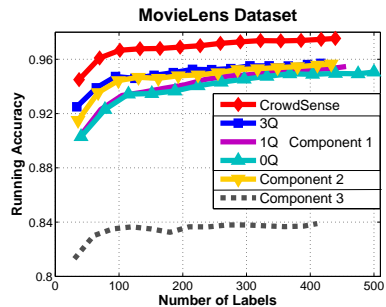}
\caption{Performance of CrowdSense and several of its variants on the Movielens dataset, averaged over 100 runs. The x-axis is the total number of votes (the total cost), and the y-axis is the running accuracy on those examples.}
\label{fig:ourbaselines}
\end{figure}
Figure \ref{fig:ourbaselines} shows an example for a specific choice of $\varepsilon$, where all three variants have lower predictive performance than CrowdSense. 

We experimented with the second component by adding labelers randomly rather than in order of their qualities. In this case, exploitation is limited, and the algorithm again tends not to perform as well (as shown in Figure \ref{fig:ourbaselines} for the curve marked ``Component 2"). 

To test the effect of the weighting scheme in the third component, we removed the use of weights from the algorithm. This corresponds to selecting the initial seed of labelers and the additional labelers at random without using their quality estimates. In addition, when combining the votes of the individual labelers, we use majority voting, rather than a weighted majority vote. This approach performed dramatically worse than the rest of the variants in Figure \ref{fig:ourbaselines}, demonstrating the significance of using quality estimates for labeler selection and the calculation of weighted vote.

\subsection{Initialization with Gold Standard}
Gold standard examples are the ``actual" opinion of the crowd gathered by taking the majority vote of all members. Even though collecting votes from the entire crowd for some of the examples initially increases the overall cost, it might help us to make better estimates in earlier iterations and thus, achieve higher overall accuracy. In CrowdSense, initialization with gold standard refers to updating step 3 in Figure \ref{fig:pseudo_crowdsense} by initializing $c_{i1}$ with the number of gold standard examples, and $a_{i1}$ with the number of times labeler $l_i$ agreed with the entire crowd. This initialization scheme is consistent with equation (\ref{eq:labeler_quality}). 

\begin{figure}[t!]
\centering
\includegraphics[width=0.65\linewidth]{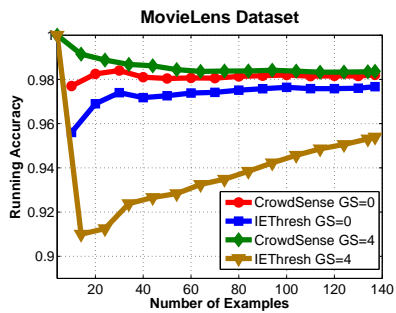}
\caption{Comparison of running accuracy with and without gold standard, averaged over 100 runs. For CrowdSense, we used $\varepsilon=0.1$ and for IEThresh, we used $\varepsilon=0.97$ to have comparable number of labelers.}
\label{fig:varying_gs}
\vspace{-5mm}
\end{figure}

In Figure \ref{fig:varying_gs}, we present the effect of initialization with gold standard for CrowdSense and IEThresh. The gold set contains four examples, where two of them were voted as +1 by the majority of the crowd, and the other two voted as -1. The running accuracy curves in Figure \ref{fig:varying_gs} start from right after observing the gold data.

As expected, gold data clearly improve CrowdSense's performance because the estimates are now initialized with perfect information for the first few examples. This benefit comes at the expense of using more of the budget at the initialization stage.

An interesting observation is that providing gold standard data to IEThresh can actually make its performance substantially \textit{worse}. Consider a labeler who agrees with the crowd's vote on every gold standard example. In this case, the labeler will get a reward for every single vote, yielding $UI=1$ (since $s(a)=0$ and $m(a)=1$). On the other hand, the labelers that  disagree with the crowd on some examples will have $s(a)>0$ and will receive $UI>1$. Therefore they will be preferred over the labelers that were in total agreement with the entire crowd's vote for each gold standard example. This illustrates why we believe that upper confidence intervals may not be the right way to handle this particular problem. In any case, we note that it may be possible to achieve better performance for IEThresh by using the upper confidence interval for the binomial distribution in early iterations rather than the $t$-distribution, since the number of correct votes is approximately binomially distributed rather than normally distributed.

\begin{figure}[b!]
\centering
\includegraphics[width=0.7\linewidth]{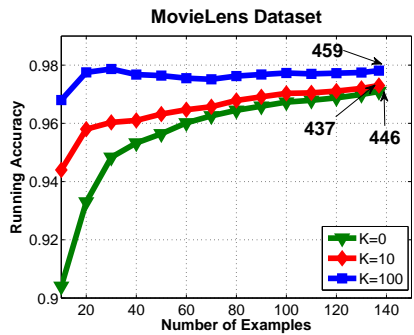}
\caption{Comparison of running accuracy of CrowdSense with varying K. All three curves use $\varepsilon=0.005$, and are averaged over 100 runs.}
\label{fig:varying_K}
\end{figure}

\begin{figure*}[t!]
\centering
\hspace{-7mm}
\subfigure {
\includegraphics[width=0.265\linewidth]{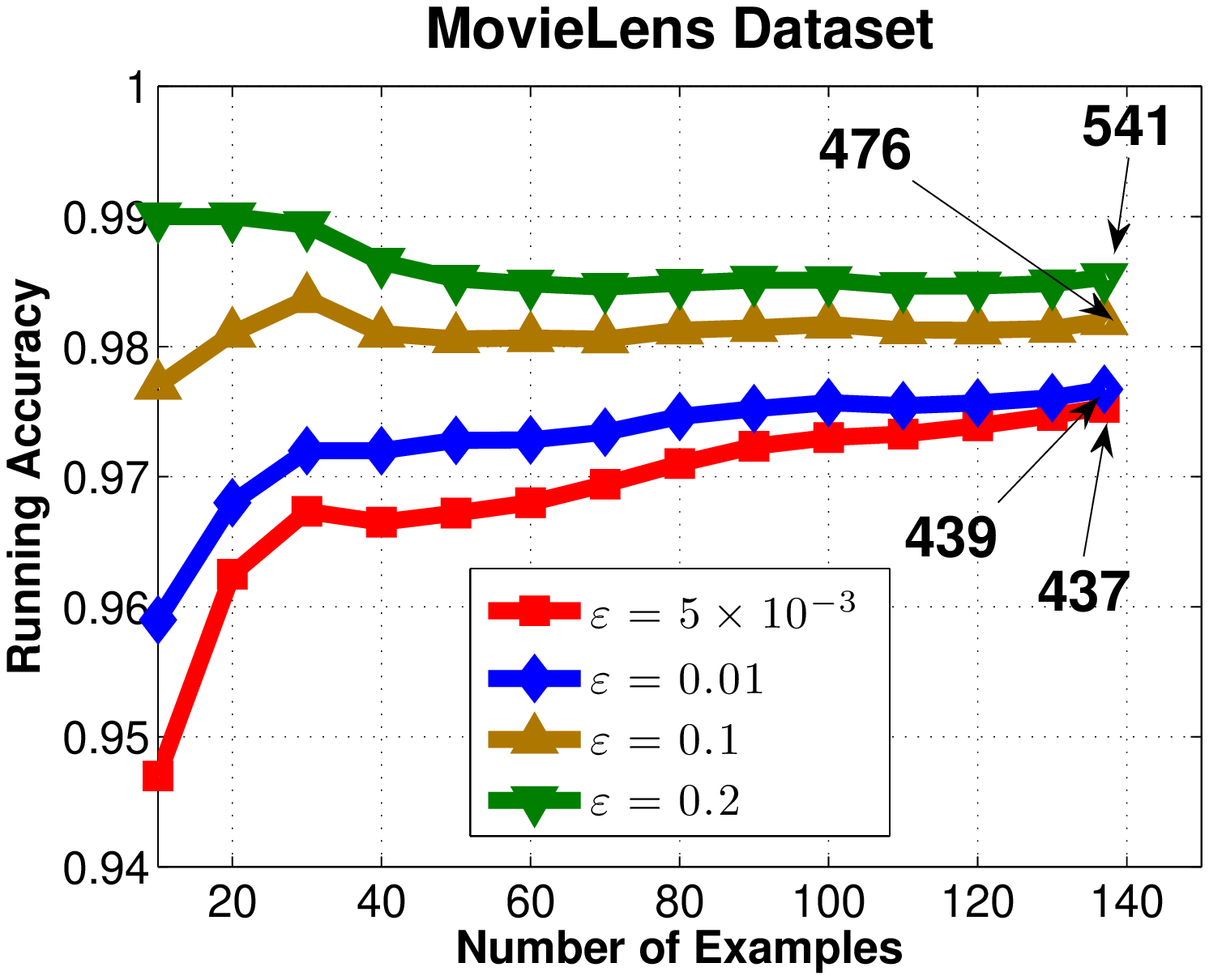}
}
\hspace{-7mm}
\subfigure {
\includegraphics[width=0.265\linewidth]{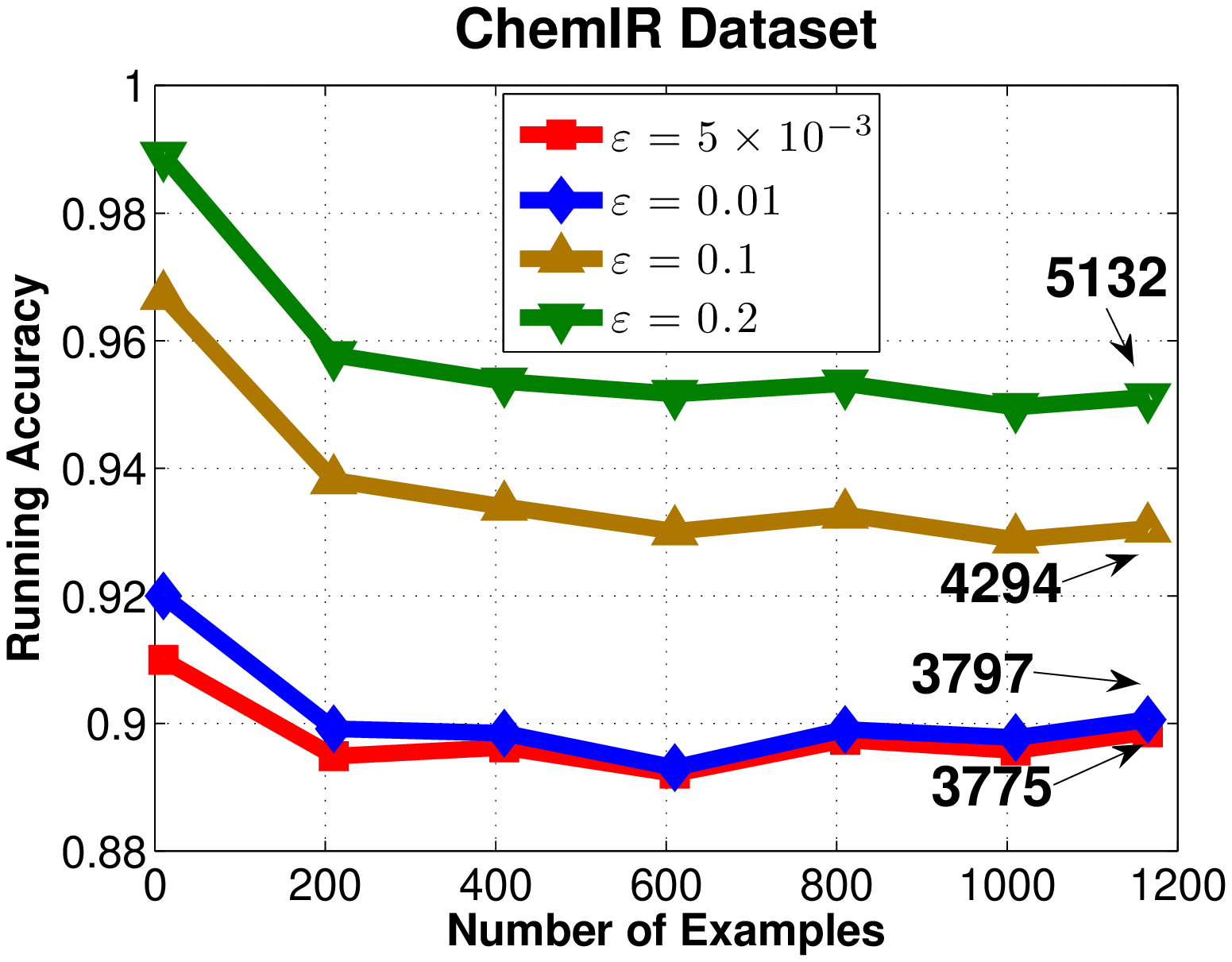}
}
\hspace{-7mm}
\subfigure {
\includegraphics[width=0.265\linewidth]{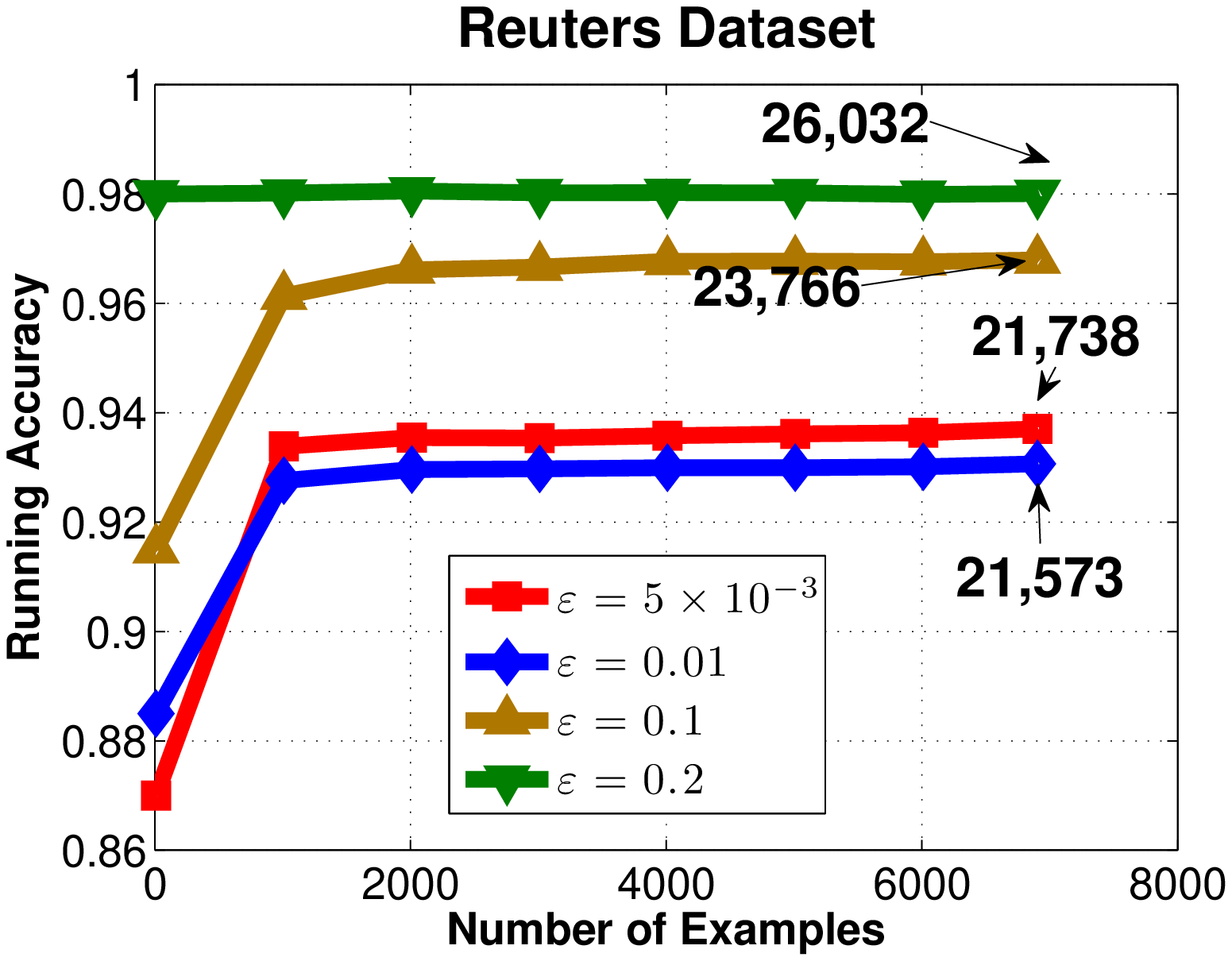}
}
\hspace{-7mm}
\subfigure {
\includegraphics[width=0.265\linewidth]{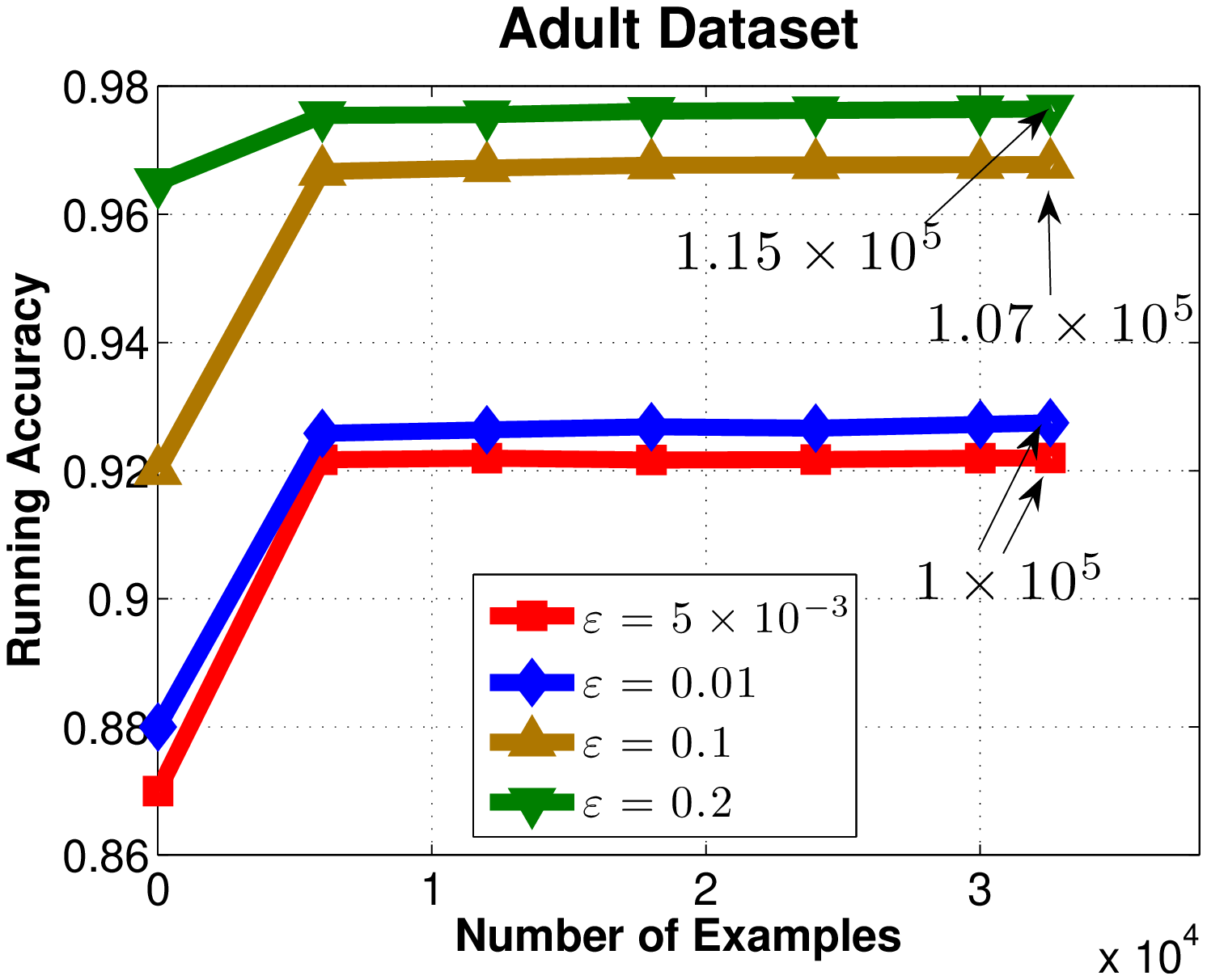}
}
\caption{Comparison of the running accuracy of CrowdSense at various $\varepsilon$ values, averaged over 100 runs. In the plots, we show how many total votes were requested at each $\varepsilon$ value.}
\label{fig:epsilon_variation}
\end{figure*}

\subsection{Effect of the $\varepsilon$ parameter}
As discussed earlier, the $\varepsilon$ parameter is a trade-off between the total cost that we are willing to spend and the accuracy that we would like to achieve. Figure \ref{fig:epsilon_variation} illustrates the average running performance over 100 runs of CrowdSense for various values of epsilon. Each final point on each of the Figure \ref{fig:epsilon_variation} curves corresponds to a single point in Figure \ref{fig:tradeoff}. Figure \ref{fig:epsilon_variation} shows that over the full time course, increasing epsilon leads to increased accuracy, at a higher cost.

\subsection{Effect of the $K$ parameter}
The $K$ parameter helps with both exploration at early stages and exploitation at later stages. The $K$ parameter helps to ensure that labelers who have seen fewer examples (smaller $c_{it}$) are not considered more valuable than good labelers who have seen many examples. The quality estimate is a shrinkage estimator, lowering probabilities when there is uncertainty. Consider a labeler who was asked to vote on just one example and correctly voted. His vote should not be counted as highly as a voter who has seen 100 examples and correctly labeled 99 of them. This can be achieved using $K$ sufficiently large.

Increasing $K$ also makes the quality estimates more stable, which helps to permit exploration. The quality estimates in the first few iterations that the labeler is involved in are not often accurate, and not very stable, and as a result, the algorithm's accuracy could be sensitive to what happens in the early iterations. Consider two equally accurate labelers. After each has labeled many items, it would have been clear that they are equal. If one labeler coincidentally makes more mistakes in the earlier iterations than the other, then the mistakes at these early iterations will constitute a larger fraction of the votes from that labeler ($a_{it}/c_{it}$ will be small, and $c_{it}$ will be small). Because of this, it is possible that this labeler will not be chosen as often as the other one, and it may not be possible to recover from the bias caused by the mistakes in the early stages. One of the purposes of $K$ is to help reduce this bias -- both labelers will be assigned almost equal quality estimates in early stages.


Since the quality estimates are all approximately equal in early stages, the weighted majority vote becomes almost a simple majority vote. This prevents CrowdSense from trusting any labeler too much early on. Having the $Q_{it}$'s be almost equal also increases the chance to put CrowdSense into the ``regime of uncertainty" where it requests more votes per example, allowing it to explore the labelers more. The impact of $K$ on the quality estimates gradually reduces as more votes are observed, and exploitation-based selection of labelers will then start favoring labelers that are indeed higher quality than others, and trusting them more in the weighted majority vote. 


We demonstrate the impact of $K$ by comparing separate runs of CrowdSense with different $K$ in Figure \ref{fig:varying_K}. At $K=100$, the quality estimates tilt the selection criteria towards an exploratory scheme in most of the iterations. This results in achieving the highest accuracy by collecting the most votes. At the other extreme, $K=0$ causes the quality estimates to be highly sensitive to the accuracy of the labels right from the start, and it can be seen that this significantly degrades the overall accuracy of the algorithm. Furthermore, it is also interesting to note that removing $K$ achieves worse performance than $K=10$ \textit{despite} collecting more votes. This indicates that the results are better when more conservative quality estimates are used in early iterations.

\section{Conclusions and Future Work}
Our goal is to ``approximate the crowd," that is, to estimate the crowd's majority vote by asking only certain members of it to vote. We discussed exploration/exploi\-tation in this context, specifically, that exploration is necessary for estimating the qualities of the labelers, and exploitation (that is, repeatedly using the best labelers) is necessary for obtaining a good estimate of the crowd's majority vote. We presented a modular outline that CrowdSense follows, which is that a small pool of labelers vote initially, then labelers are added incrementally until the estimate of the majority is more certain, then a weighted majority vote is taken as the overall prediction. We discussed specific choices within this modular outline, the most important one being the choice of $\varepsilon$, which determines the overall budget for labeling the entire dataset. We compared our results to several baselines, indicating that CrowdSense, and the overall exploration/exploitation ideas behind it, can be useful for approximating the crowd.

One of the main challenges to approximating the crowd has to do with the fact that the majority vote is taken as the ground truth (the truth we aim to predict). This means that there is a complicated relationship (a joint probability distribution) between the labelers' accuracies and the ground truth. In the longer version of this work, we propose variations of CrowdSense that directly incorporate assumptions about this joint distribution. In particular, we consider ($i$) a statistical independence assumption of the labelers, which is useful for modeling large crowds, and ($ii$) a lower bound on how often the current sub-crowd agrees with the crowd majority vote, where calculations are based on the binomial distribution. In both of these cases, even though probabilistic assumptions were made, the accuracy of the resulting algorithm is comparable to (or lower than) CrowdSense itself. It is difficult to characterize the joint distribution for the problem of approximating the crowd, due to the constraint that the majority vote is the true label. CrowdSense, with its easy-to-understand weighted majority voting scheme, seems to capture the essence of the problem, and yet has the best performance within the pool of algorithms we tried.

\section{Acknowledgments}
We would like to thank Thomas W. Malone for helpful discussions. Support for this project is provided by MIT Intelligence Initiative (I$^2$).

\bibliography{references}  
\balancecolumns
\end{document}